# BACKGROUND ISSUES AT SUPERDAΦNE

M. Boscolo, LNF, Frascati, Italy

*Abstract*

DAΦNE induced backgrounds are dominated by Touschek scattering [1]. A luminosity upgrade of the machine at the same beam energy would be obtained essentially by increasing the bunch density, that in principle would increase this particle loss process.

The beam pipe along the ring and especially at the Interaction Region (IR) has to be designed to minimize losses as much as possible. A masking system has to be incorporated to shield the detector from beam-generated background. Collimators are required upstream the Interaction Point (IP) to remove errant beam particles away from the detector. Shape and position of collimators and masks at the IR must be optimized with simulations. Also scrapers along the ring have to be carefully planned.

All these studies can be performed using the same tool used for the DAΦNE case [2]. In fact, the reliability of this program has been tested with the KLOE data, showing a good agreement with simulations.

Some preliminary studies and general considerations of the Touschek effect for the DAΦNE high luminosity upgrade are presented here.

## TOUSCHEK SIMULATION BASIC IDEAS

Touschek scattering is a source of off-energy beam particles arising from the elastic scattering of particles within a bunch. Scattering results in two particles with energy errors $+\delta p/p$ and $-\delta p/p$ which follow betatron trajectories around the off-energy closed orbit.

A simulation code STAR has been developed for background studies at DAΦNE. The simulation code uses the MAD output. It performs an optics check, then it calculates the synchrotron integrals from which some beam parameters (i.e.: emittance, energy spread and momentum compaction) are derived. The Touschek energy spectra are evaluated using these values.

Touschek particles are taken within one transversely Gaussian bunch with the proper energy spectra. Particles are tracked over many turns or until they are lost. In this way an estimation of the Touschek losses along the whole ring and at the IRs is performed.

### General Considerations

- Most of the losses arise from the Touschek scattered particles in dispersive regions, so only these particles are simulated. Touschek scattered particles have a betatron oscillation which is proportional to the dispersion D, to the invariant H ($\propto$D) and to the momentum spread $\Delta p/p$:

$$x = \frac{\delta p}{p}(|D| + \sqrt{H\beta}).$$

The parameter H-invariant is defined by the following relation:

$$H = \gamma_x D_x^2 + 2\alpha_x D_x D_x' + \beta_x {D_x'}^2.$$

- With a given energy spectrum P(E) one can either extract the single scattered particle energy shift according to it or use a uniform extraction and use P(E) as a weight. We use the second possibility. In this way we cope with the tails of both the Touschek probability density function and the probability of beam loss versus energy deviation. The behaviour of these two functions versus the beam energy shift is opposite one to the other. In fact, for a lower energy shift, the Touschek scattering probability increases while the probability of loss decreases. The Touschek density function is mostly related to the beam parameters like bunch volume, emittance, momentum deviation and bunch current. On the other hand, particle losses are related mostly to machine parameters and optics, like the physical aperture, the phase advance between dispersive regions and collimators and between dispersive regions and IR.

The calculation of the energy spectra is done starting from the formula [3]:

$$\frac{1}{\tau} = \frac{\sqrt{\pi} r_e^2 cN}{\gamma^3 (4\pi)^{3/2} V \sigma_x' \varepsilon^2} C(u_{min})$$

where: $\varepsilon = \frac{\Delta E}{E}$, $u_{min} = \left(\frac{\varepsilon}{\gamma \sigma_x'}\right)^2$, $V = \sigma_x \sigma_y \sigma_l$

and $\sigma_x' = \sqrt{\frac{\varepsilon_x}{\beta_x} + \sigma_p^2 (D_x' + D_x \frac{\alpha_x}{\beta_x})^2}$.

$C(u_{min})$ accounts for the Moller cross-section and momentum distribution. The presence of transverse beam polarization can be taken into account.

For a chosen machine section the Touschek probability is evaluated in small steps (9 per element) to account for the beam parameters evolution. This is repeated for 100 $\varepsilon$ values. An interpolation between the calculated $\varepsilon$ values according to the Touschek scaling law [1] $A_1 \varepsilon^{-A_2}$ allows to find the density function for the chosen section.

# MAIN BACKGROUND STUDIES AT DAϕNE

The simulation code STAR is used to calculate the locations where the off-energy particles hit the vacuum chamber of the ring, with particular care at the position of the losses at the two interaction regions (IRs) [2]. The particles are tracked over many turns and those lost at the KLOE IR have been fully simulated in the detector allowing the evaluation of the background counting rates and of detailed studies of its properties, namely spatial distribution and energy spectra. Several comparisons have been performed between simulations and KLOE data showing a good qualitative agreement [4].

The program has been used to optimize shape and position of additional scrapers added in dispersive regions. It has also been useful to estimate the expected variation of background rates at different optics changes. During the last DAΦNE shut-down (first months of year 2003) additional masks have been studied and implemented to shield the KLOE focusing quadrupoles, which became stronger in the IR doublet design.

## SUPERDAϕNE LATTICE

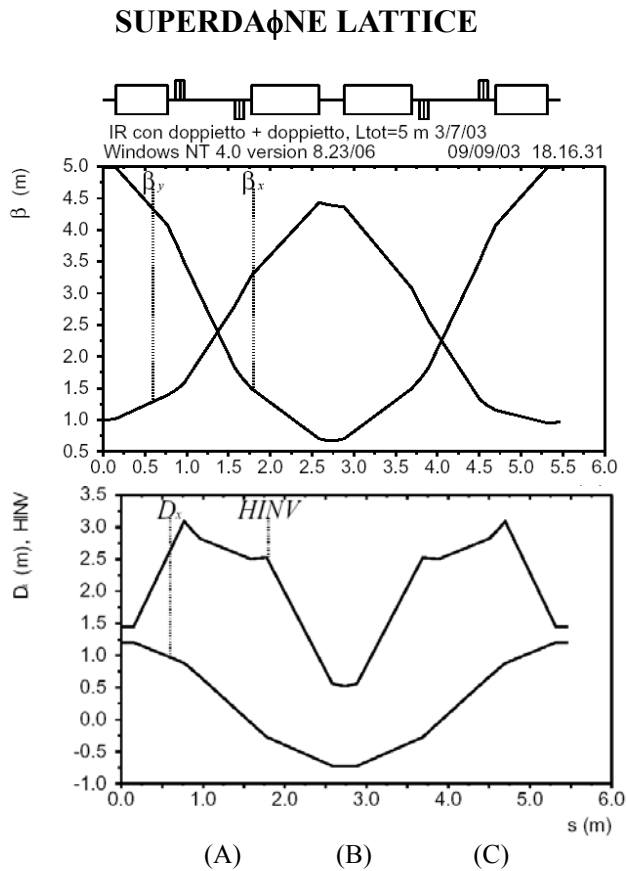

(A)   (B)   (C)

Fig.1 One cell lattice functions. *Upper plot*: $\beta_x$ and $\beta_y$; *lower plot*: horizontal dispersion and H-invariant.

The machine has been designed [5] with two straight sections: one for the interaction region, the other for the RF cavity. In between there are 6 long cells on one side and 5 short cells on the other. Each cell has a phase advance of π. So, total losses can be obtained multiplying the contributions of one cell by the number of cells.

Fig. 1 shows the lattice functions for one cell: upper plot shows the β-functions, lower plot dispersion and H-invariant. Each cell has 3 regions (A, B, C) with an essentially constant H value, defined in fig.1. The Touschek scattered particles gain high betatron oscillations in the regions with high H and D values, that are regions (A) and (C) of fig.1. The scattered particles of these two regions have equal energy spectra but different phase advance, so the losses arising from the two locations may result very different.

The Touschek probability density function is plotted in fig.2 for the constant H value regions. Regions A and C have equal weights (higher blue curve), region B has lower value (lower red line). In the plot $10\sigma_p$ is at 0.01.

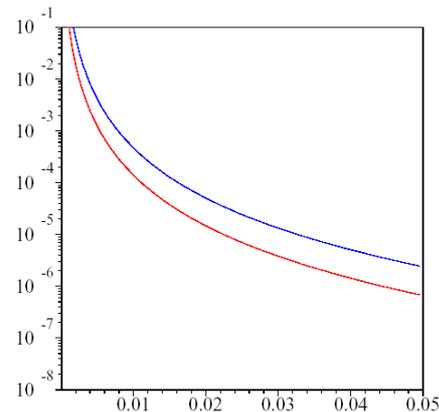

Fig.2 Touschek probability density function as a function of the energy deviation within one cell: upper blue line is for regions A and C, red curve for region B.

## TRACKING

The trajectories of the Touschek particles and their hit positions along the beam pipe have been studied with the simulation all along the ring. A physical aperture of 5 cm has been taken for the beam pipe all over the ring but at the IR. In fact, as a first attempt, the IR physical aperture has been taken at $20\sigma_x$ of the beam.

The beam parameters used for these simulations are reported in table 1.

The calculated trajectories of the particles scattered in the cell upstream the IR are shown in fig. 3 and 4. In particular, Touschek scattered particles are generated separately in the 3 regions of the cell. Calculated losses of scattered particles starting from region B are one order of magnitude lower than those coming from regions A and C. Upper plot of fig.3 shows the trajectories of particles scattered in region C of the cell upstream the IR; lower plot those generated in region A. The radial phase difference of the two families of particles is evident. For the upper case particles get lost right at the IR, while for the lower one the phase is such that particles get lost before they arrive at the IR. As a consequence, for this latter case calculated losses at the IR result half of the upper case. Moreover, lower plot of fig. 3 shows that Touschek trajectories are quite large upstream the IR, indicating that collimators should be inserted upstream the IR. In particular, if they were at about -5m far from the IP they would stop all scattered particles (see lower plot of fig. 3). As a general consideration, to cut scattered particles with all the different phases the collimators should be at least two at 90º phase.

Table 1. Beam parameter set used for simulations.

| | |
|---|---|
| $E_{beam}$ [MeV] | 510 |
| $I_{bunch}$ [mA] | 20 |
| $\varepsilon_x$ [m rad] | $0.3 \cdot 10^{-6}$ |
| $\kappa$ | 0.01 |
| $<\sigma_l>$ [cm] | 1 |
| $\sigma_p$ | $10^{-3}$ |
| $\alpha_c$ | -0.23 |
| H | 175 |
| $V_{RF}$ [MV] | 8.5 |
| $\beta_x$ (@IP) [m] | 0.5 |
| $\beta_y$ (@IP) [mm] | 4 |
| $\sigma_z$ (@IP) [mm] | 3.7 |
| L [m] | 100 |

In fig.4 trajectories are tracked along the ring. As appears from the plot, particles never get lost along the ring, indicating that the chosen physical aperture of 5cm is enough.

Total loss rates have been estimated and compared to the DAΦNE case -with the optics used for the KLOE runs summer 2002. Rates are reported in table 2 and are referred without scrapers. Of course, the DAΦNE case is the optimized one, with the real physical aperture. On the other hand, for the DAΦNE upgrade the calculated rates are a preliminary number indicating that a beam pipe of 5cm is enough along the ring. Moreover, as losses are concentrated at the IR, the indication is that a physical aperture of 20 $\sigma_x$ of the beam at the IR is not enough. A detailed optimization of the IR design must be performed, also to minimize the Touschek particles losses. As shown in fig. 5 a 30% increase of the IR aperture reduces calculated losses by 50%.

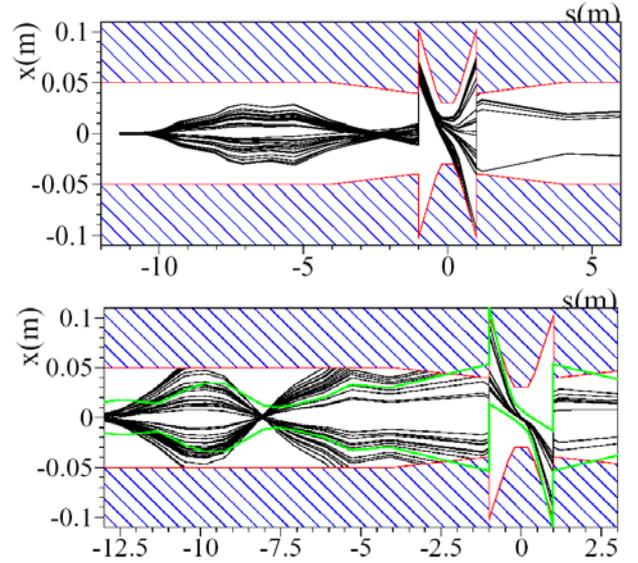

Fig.3. Trajectories of particles starting from the cell upstream the IR and tracked through the IR. Upper plot: generation in region C; lower plot: generation in region A and $20\sigma_x$ (green line).

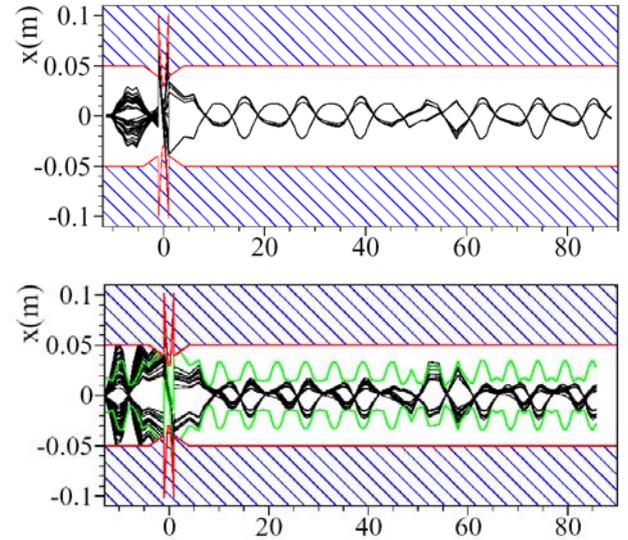

Fig.4. Trajectories of particles (black lines) starting from the cell upstream the IR and tracked all along the ring, with a physical aperture of 5cm everywhere but in the IR. Upper plot: generation in region C; lower plot: generation in region A and $20\sigma_x$ (green line).

Table 2. First evaluations of Touschek loss rates.

| KHz (@10mA) | SUPERDAΦNE | DAΦNE |
|---|---|---|
| Total Rates | 2500 | 1800 |

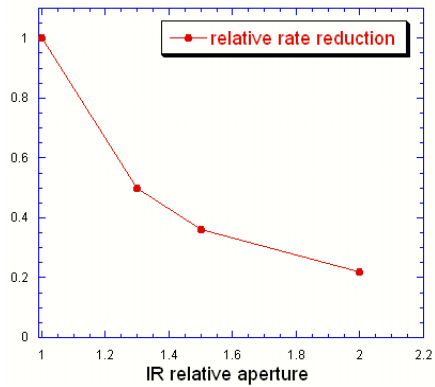

Fig.5. Relative loss rate reduction as a function of a relative increase of the IR physical aperture.

## CONCLUSIONS

The Touschek simulations have been successfully used at DAΦNE. The same tool can be used for the SUPERDAΦNE design in order to define the position and shape of collimators and masks, to design the beam pipe in the ring, especially at the IR and to optimize the horizontal phase advance between last cell and the IP.